\documentclass[prb,showpacs,amsmath,amssymb,superscriptaddress,twocolumn,letterpaper]{revtex4}

\usepackage{amssymb}
\usepackage{graphicx}
\usepackage{color}
\usepackage{dcolumn}
\usepackage{bm}

\begin{document}

\title{High pressure transport studies of the LiFeAs analogues CuFeTe$_2$ and Fe$_2$As}

\author{D. A. Zocco}
\altaffiliation[Present address: ]{Institute of Solid State Physics (IFP), Karlsruhe Institute of Technology, D-76021 Karlsruhe, Germany.}
\author{D. Y. T\"{u}t\"{u}n}
\author{J. J. Hamlin}
\affiliation{Department of Physics, University of California, San Diego, La Jolla, California 92093, USA}
\author{J. R. Jeffries}
\affiliation{Condensed Matter and Materials Division, Lawrence Livermore National Laboratory, Livermore, California 94550, USA}
\author{S. T. Weir}
\affiliation{Condensed Matter and Materials Division, Lawrence Livermore National Laboratory, Livermore, California 94550, USA}
\author{Y. K. Vohra}
\affiliation{Department of Physics, University of Alabama at Birmingham, Birmingham, Alabama 35294, USA}
\author{M. B. Maple}
\email[E-mail: ]{mbmaple@ucsd.edu}
\affiliation{Department of Physics, University of California, San Diego, La Jolla, California 92093, USA}

\begin{abstract}
We have synthesized two iron-pnictide/chalcogenide materials, CuFeTe$_2$ and Fe$_2$As, which share crystallographic features with known iron-based superconductors, and carried out high-pressure electrical resistivity measurements on these materials to pressures in excess of 30 GPa. Both compounds crystallize in the Cu$_2$Sb-type crystal structure that is characteristic of LiFeAs (with CuFeTe$_2$ exhibiting a disordered variant). At ambient pressure, CuFeTe$_2$ is a semiconductor and has been suggested to exhibit a spin-density-wave transition, while Fe$_2$As is a metallic antiferromagnet. The electrical resistivity of CuFeTe$_2$, measured at 4 K, decreases by almost two orders of magnitude between ambient pressure and 2.4 GPa. At 34 GPa, the electrical resistivity decreases upon cooling the sample below 150 K, suggesting the proximity of the compound to a metal-insulator transition. Neither CuFeTe$_2$ nor Fe$_2$As superconduct above 1.1 K throughout the measured pressure range.
\end{abstract}


\maketitle

The recent discovery of the Fe-based superconducting materials has extended the field of high-temperature superconductivity beyond the thoroughly studied copper-oxides. However, the older cuprates still retain the record for the highest superconducting transition temperatures $T_c$ (166 K at a pressure of $\sim 300$ kbar, Refs.~\onlinecite{mao_1994_1,miguelrecord}) while in the newer Fe-based superconductors, maximum $T_c$ values of 55 K have been achieved in the ``1111'' compound SmFeAsO substituted with fluorine.\cite{ren_2008_1} Lower values of $T_c$ were found in other sub-families of Fe-based superconductors, the ``11,'' ``111,'' ``122,'' and ``21311'' compounds (for reviews, see, for example, Refs.~\onlinecite{paglione10,Johnston2010a,stewart2011a}). All these sub-families of Fe-based superconductors share the same building blocks, namely tetrahedra composed of pnictogen or chalcogen atoms surrounding an Fe atom at the center, whose shape seems to play an important role in determining the maximum $T_c$ achievable in any particular sub-family of compounds.\cite{wilson_2010_1} An important question that arises is if there exist other classes of compounds with similar building blocks where superconductivity develops at comparable or higher values of $T_c$.
\begin{figure}[htb]
\begin{center}
{\includegraphics[width=2.5in]{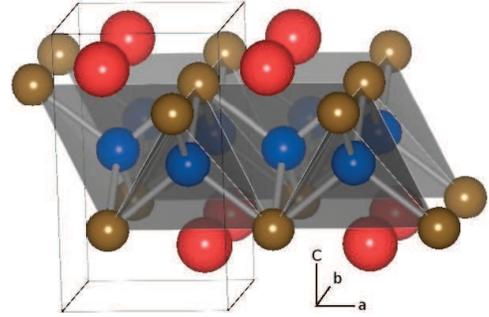}}
\caption{Cu$_2$Sb structure type exhibited by LiFeAs, CuFeTe$_2$, and Fe$_2$As.  Blue, brown, and red atoms occupy the 2(a), 2(c$_{\mathrm{I}}$), and 2(c$_{\mathrm{II}}$) sites, respectively.  For LiFeAs, blue is Fe, brown is As, and red is Li.  For Fe$_2$As, blue and red are Fe, and brown is As.  In the case of CuFeTe$_2$, brown is Te, while blue and red represent an approximately equal mixture of Cu and Fe.  CuFeTe$_2$ adopts a defect variant of the Cu$_2$Sb structure, where the 2(c$_{\mathrm{II}}$) (red) sites are only partially occupied (occupancy $\lesssim15\%$).} \label{fig1}
\end{center}
\end{figure}

Recently, it was found that superconductivity appears upon the application of pressure to a non-magnetic analog of the Fe-based superconductors, SnO (Ref.~\onlinecite{forthaus10}). The compound SnO is a diamagnetic semiconductor at ambient pressure; it forms in the $\alpha$-PbO structure adopted by $\beta$-FeSe, which enters the superconducting state below $T_c$ = 8 K at ambient pressure.\cite{chu_2009_1} Above a pressure $P \sim 5$ GPa, SnO becomes metallic at room temperature, while the crystal structure remains unchanged up to $P$ $\sim$ 17 GPa. In that work, the authors showed that SnO becomes superconducting under pressure, where $T_c(P)$ has a dome-like shape, with a maximum $T_c$ of 1.4 K at $P$ $\sim$ 9 GPa, similar to other Fe-based superconductors. Following the same line of thought, we have embarked on a search for pressure-induced superconductivity in materials that contain Fe and crystallize with similar layers of Fe-$Pn/Ch$ ($Pn$ = pnictogen and $Ch$ = chalcogen) tetrahedra as found in known iron-based superconductors.

\begin{table}[h]
\caption{Structural parameters of the Cu$_2$Sb-type compounds CuFeTe$_2$ \onlinecite{lamarche98} and Fe$_2$As \onlinecite{pearson_1985_1} compared with those of LiFeAs \onlinecite{lynn_2009_1}.}
\begin{center}
\begin{tabular}{l|l|l|l|l|l|l|l|l}
× & $a$ (\AA{}) & $c$ (\AA{}) & $c/a$ & $2(a)$ & $2(c_{\mathrm{I}})$ & $z_{\mathrm{I}}$ & $2(c_{\mathrm{II}})$ & $z_{\mathrm{II}}$\\
\hline
LiFeAs & 3.78 & 6.35 & 1.68 & Fe & As & 0.26 & Li & 0.85\\
CuFeTe$_2$ & 3.98 & 6.08 & 1.53 & Fe/Cu & Te & 0.279 & Fe/Cu & 0.72\\
Fe$_2$As & 3.63 & 5.99 & 1.65 & Fe & As & 0.265 & Fe & 0.67
\end{tabular}
\end{center}
\label{table1}
\end{table}

Two compounds containing layers of Fe-$Pn/Ch$ tetrahedra are CuFeTe$_2$ and Fe$_2$As, which both adopt the Cu$_2$Sb structure (space group P4/nmm) that is also found for LiFeAs ($T_c \sim 18$ K).\cite{wang_2008_1,Pitcher_2008_1}  This structure is characterized by tetrahedrally bonded layers interspersed with additional atoms in between the layers (see Fig.~\ref{fig1}).  In the case of LiFeAs, these additional atoms are Li, for CuFeTe$_2$, they are a disordered mixture of Fe and Cu, while for Fe$_2$As, they are Fe atoms.  In the case of CuFeTe$_2$, the interlayer sites at $2(c_{\mathrm{II}})$ (0, $\frac{1}{2}$, $z_{\mathrm{II}}$) are only partially filled with Cu and Fe atoms, with an occupancy of less than 15$\%$ (Ref.~\onlinecite{lamarche98}). For this reason, the structure of CuFeTe$_2$ is closely related to the PbO structure, which lacks additional interlayer atoms.  The structural parameters of LiFeAs, CuFeTe$_2$, and Fe$_2$As are compared in Table~\ref{table1}.  The compound CuFeTe$_2$ is a semiconductor at ambient pressure,\cite{vaipolin92} and it displays rather ambiguous magnetic properties: paramagnetism between 10 and 200 K was observed in single crystals,\cite{vaipolin92} spin-glass behavior was reported for polycrystalline samples,\cite{lamarche98} and, more recently, spin-density wave (SDW) ordering near room temperature was proposed.\cite{rivas98,gonzalezjimenez99,rivas01} In contrast, Fe$_2$As is metallic at ambient pressure with a well established antiferromagnetic transition near 350 K.\cite{katsuraki_1966_2}

In this paper, we report the synthesis of CuFeTe$_2$ single crystals and Fe$_2$As polycrystals and present the results of measurements of their transport properties under high pressure conditions. High pressure electrical resistivity measurements on CuFeTe$_2$, which display non-metallic behavior at low pressures, suggest a possible insulator to metal transition in the vicinity of 34 GPa. Preliminary high pressure electrical resistivity measurements on Fe$_2$As to 30 GPa, indicate that the transport properties of this material are rather insensitive to pressure.

\section{Experimental details}
\begin{figure}
\begin{center}
\includegraphics[width=3in]{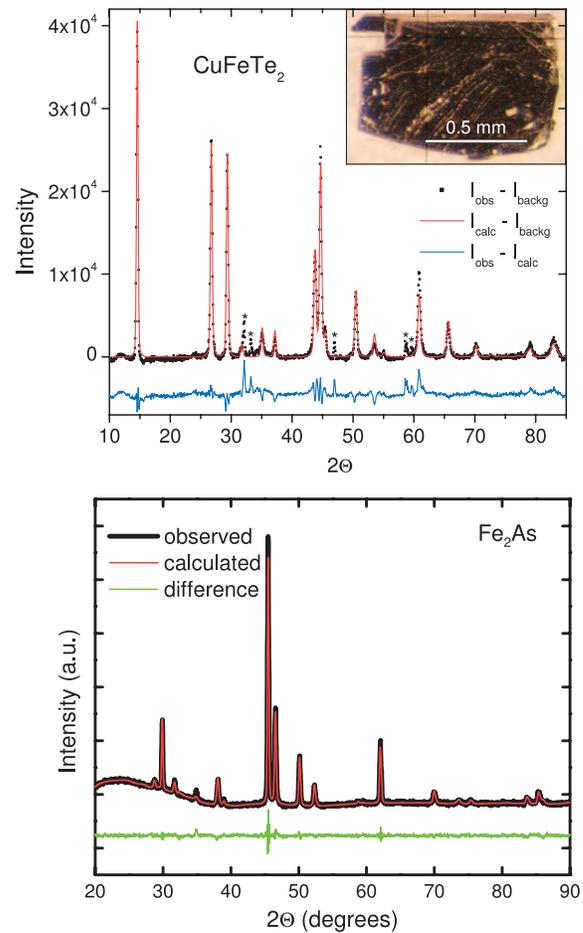}
\caption{Powder x-ray diffraction patterns of CuFeTe$_2$ (left) and Fe$_2$As (right)  Fits to the data were obtained via Rietveld analysis.  Additional reflections in the CuFeTe$_2$ data, indicated by `$\ast$', were identified as impurity peaks due to FeTe$_{2}$ inclusions. \textit{Inset}: Photograph of a single crystal of CuFeTe$_2$. Typical platelets are $\sim$ 50 $\mu$m thick.} \label{fig2}
\end{center}
\end{figure}

In a previous work,\cite{lamarche98} it was found that samples of CuFeTe$_2$ prepared with the stoichiometric ratio of elements were not single phase. For this reason, single crystals with nominal composition of Cu$_{1.13}$Fe$_{1.22}$Te$_2$ were grown as described in Ref.~\onlinecite{vaipolin92} (we refer to the samples as ``CuFeTe$_2$'' for simplicity). Samples were prepared from the melt, using high purity Cu shot, Fe powder (22 mesh), and a piece of Te ingot. Elements were weighed inside an argon filled glove-box and sealed under vacuum in a quartz tube. The tube was heated for 24 hours at $1100^{\circ}$C, slowly cooled to $600^{\circ}$C at a rate of $50^{\circ}$C/h, and allowed to dwell at $600^{\circ}$C for another 100 h, after which the furnace was turned off and allowed to cool to room temperature. The $inset$ in Fig.~\ref{fig2} shows a photograph of a single crystal that was separated mechanically from the main boule. The samples are 1 mm gold colored platelets, of micaceous character, which make them easy to cleave along the $ab$-planes.

For the synthesis of Fe$_2$As, a mixture of iron powder (99.99\%) and arsenic chunks (99.99\%) were combined in the stoichiometric ratio in a alumina crucible and sealed in an evacuated quartz tube that was heated slowly from room temperature to $600^{\circ}$C. Due to the high vapor pressure and reactivity of As, the sample was kept at $600^{\circ}$C for 10 hours in order to pre-react the sample and avoid the excessive pressure inside the quartz tube that could develop by rapid heating. The sample was then heated to $1060^{\circ}$C over a period of one day, followed by a slow cool to room temperature.

X-ray powder diffraction (XRD) measurements were made at room temperature using a Bruker D8 Discover diffractometer utilizing CuK$_{\alpha}$ radiation, in order confirm the crystal structure and check the purity of the samples. Due to their malleability, the CuFeTe$_2$ crystals were difficult to grind into a fine powder, resulting in preferential orientation in the powder diffraction results. The coarse powder was obtained by grinding several crystals using a mortar and pestle.  The results of the XRD measurements are presented in Figure~\ref{fig2}.  Rietveld refinement of the XRD pattern using the General Structure Analysis System (GSAS),\cite{gsas} revealed a good agreement with the Cu$_2$Sb structures discussed above, with lattice constants consistent with those listed in Table~\ref{table1}.  The CuFeTe$_2$ samples show peaks indicating a small amount ($\sim 5\%$) of FeTe$_2$ impurity.\cite{pertlik86} For the Fe$_2$As sample, no substantial impurity is detectable from the XRD results. Magnetic susceptibility measurements on Fe$_2$As are consistent with a ferromagnetic impurity phase containing at most 0.6\% of the iron.

\begin{figure}[t]
\begin{center}
{\includegraphics[width=3.4in]{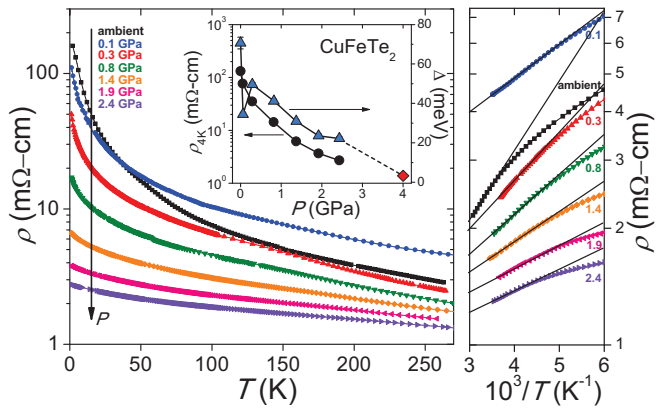}}
\caption{\textit{Left panel}: Electrical resistivity $\rho$ versus temperature $T$ of CuFeTe$_2$ obtained in the hydrostatic piston-cylinder clamped cell experiment, for increasing pressures $P$ indicated by the arrow. The $inset$ shows the values of resistivity $\rho_{4K}$ measured at 4 K (\textit{left axis}) and the energy gap values $\Delta$ (\textit{right axis}) as a function of pressure, obtained as described in the main text (\textit{blue triangles}: piston-cylinder cell; \textit{red diamond}: diamond-anvil cell). \textit{Right panel}: High-temperature fits to estimate the energy gap.} \label{fig3}
\end{center}
\end{figure}

Ambient pressure electrical resistivity and specific heat measurements were performed in a Quantum Design Physical Properties Measurement System (PPMS).  For electrical resistivity measurements, the samples were cut into bars, gold contact pads were sputtered onto each sample, and four gold wires were attached using two part silver epoxy (Epotek H20E).  For CuFeTe$_2$, the electrical resistivity was measured in the $ab$-plane.  Electrical resistivity measurements under pressure were performed employing two techniques. In the low-pressure range, hydrostatic pressures up to 2.4 GPa were applied with a beryllium-copper, piston-cylinder clamped cell using a Teflon capsule filled with a 1:1 mixture of n-pentane:isoamyl alcohol as the pressure transmitting medium to ensure hydrostatic conditions during pressurization at room temperature. The pressure in the sample chamber was inferred from the inductively determined, pressure-dependent superconducting critical temperature of a tin manometer.\cite{smith69} Pressure gradients for the hydrostatic cell ($\delta P$ $<$ 2\%) were inferred from the width of the superconducting transition of the manometer.

\begin{figure}[t]
\begin{center}
{\includegraphics[width=3.2in]{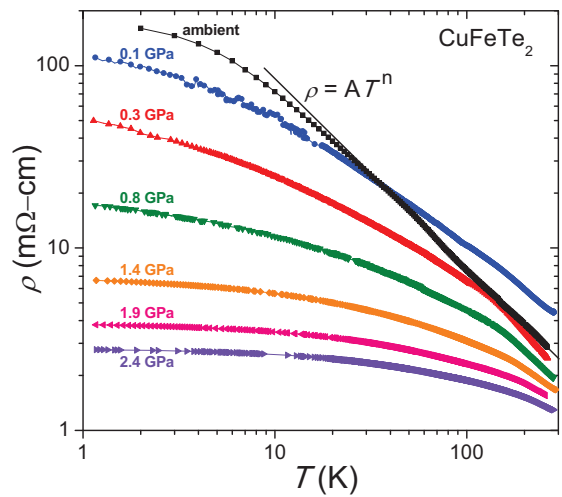}}
\caption{$\log$-$\log$ plot of the electrical resistivity $\rho$ versus temperature $T$ of CuFeTe$_2$ obtained in the piston-cylinder clamped cell experiment. A power-law fit to the ambient pressure curve for temperatures ranging from 30 K to 270 K yielded a value of the exponent $n \sim -1$.} \label{fig4}
\end{center}
\end{figure}

For higher pressures, a diamond-anvil cell (DAC) was used. The DAC is a mechanically loaded commercial model, manufactured by Kyowa Seisakusho Limited. A 4-lead ``designer'' diamond~\cite{weir00} was used with a culet size of 190 $\mu$m in diameter and distance between the electrical resistivity probes of $\sim$ 30 $\mu$m.  The gasket was made from a 200 $\mu$m thick MP35N foil preindented to 40-50 $\mu$m and a 100 $\mu$m diameter hole was drilled through the gasket using an electrical discharge machine (EDM). In order to protect the designer diamond throughout repeated use, a maximum pressure of only 34 GPa was applied. For the measurements on CuFeTe$_2$ the sample was surrounded by steatite powder, which is a soft solid that acts as a quasi-hydrostatic pressure medium, and a small chip of crystal was oriented to measure the resistance in the $ab$-plane.  Pressure was adjusted and determined at room temperature, using the fluorescence spectrum of several $\sim$ 5 $\mu$m diameter ruby spheres located in the sample space and the calibration of Chijioke \textit{et al.},\cite{chijioke05} with a $\delta P$ $\leq$ 15\%, inferred from the full width at half maximum (FWHM) of the fluorescence line. Further details of the DAC technique are described in Ref.~\onlinecite{jackson06}. For the Fe$_2$As experiments, no pressure medium was utilized, and the sample completely filled the gasket hole. In all of the high pressure measurements, the electrical resistance was measured using a 4-lead technique and a Linear Research Inc. LR-700 AC resistance bridge operating at 16 Hz. The experiments at high pressure were performed from room temperature to 1.2 K in a conventional pumped $^4$He dewar.

\section{Results and Discussion}
The left panel of Fig.~\ref{fig3} shows the high-pressure electrical resistivity $\rho$ versus temperature $T$ of CuFeTe$_2$ derived from the hydrostatic cell experiment. The ambient pressure curve was measured in the PPMS, and the same sample was used afterwards in the high pressure measurements. At ambient pressure, CuFeTe$_2$ displays non-metallic behavior throughout the entire temperature range; upon application of pressure, the resistivity is rapidly suppressed. The $inset$ of Fig.~\ref{fig3} displays the resistivity values measured at 4 K as a function of applied pressure (\textit{black circles}). Modest pressures of 2.4 GPa are sufficient to reduce the magnitude of the resistivity by almost two orders of magnitude and cause the resistivity to lose most of it temperature dependence, in comparison with the ambient pressure curve.

In the high-temperature region (150 K to room temperature), the $\rho (T)$ data exhibit ``activated'' behavior and can be described by an exponential function $\rho (T) = A\exp(\Delta/2k_{B}T)$, where $\Delta$ is the energy gap, $A$ is a constant and $k_B$ is the Boltzmann constant. The energy gap $\Delta$ was obtained by plotting $\ln \rho (T)$ versus $1/T$ and fitting a straight line to the data. The $inset$ of Fig.~\ref{fig3} displays the values of $\Delta$ as a function of pressure obtained from this procedure, which are reduced by the application of pressure (\textit{blue triangles}: piston-cylinder clamped cell; \textit{red diamond}: diamond-anvil cell). The energy gap value obtained from our electrical resistivity data at ambient pressure is $\sim$ 3 times larger than the previously reported value,\cite{abdullaev06} although the value extracted for the gap is sensitive to the temperature region selected for the fit since the data (at ambient pressure) can not be described over any appreciable temperature range by the exponential form discussed above. The non-monotonic dependence of the extracted energy gap on pressure may be related to the fact that this simple treatment of the temperature dependence of the resistivity does not capture the full physics of this material, since previous analyses have suggested that CuFeTe$_2$ may be a gapless semiconductor at ambient pressure.

\begin{figure}[htb]
\begin{center}
{\includegraphics[width=3.3in]{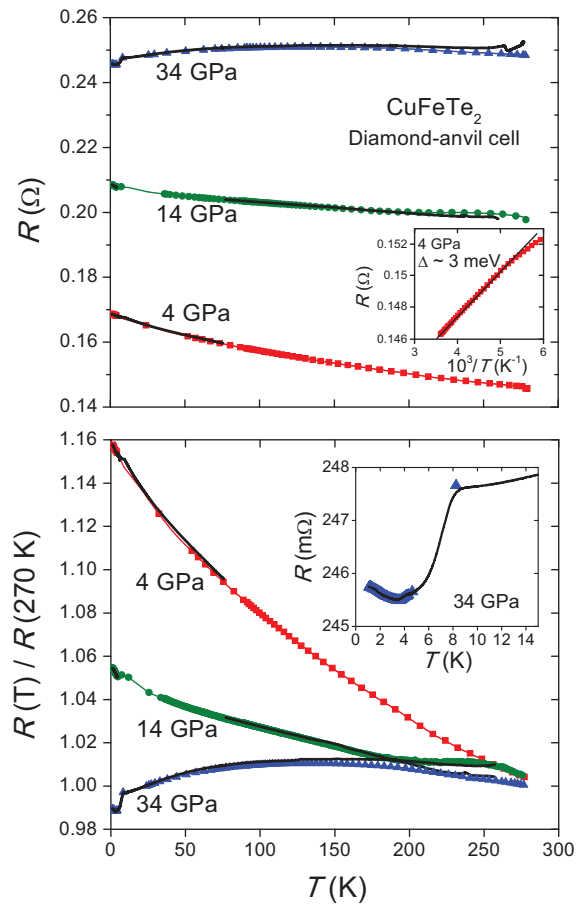}}
\caption{Electrical resistance $R$ versus temperature $T$ of CuFeTe$_2$ obtained in the diamond-anvil cell experiment. Red, green and blue curves were measured during the cooling cycles for pressures of 4, 14 and 34 GPa, respectively, while the black curves correspond to the measurements taken while warming the probe. An energy gap of $\sim$ 3 meV was obtained from the fit to the high temperature region of the measurement taken at 4 GPa (upper panel $inset$).  The lower panel displays the resistance curves normalized to their respective values of resistance measured at 270 K. The $inset$ shows in detail the low temperature region of the 34 GPa curve.} \label{fig5}
\end{center}
\end{figure}

CuFeTe$_2$ was identified as a gapless magnetic semiconductor, due to the observation of power-law, $\propto T^n$, temperature dependences of the Hall coefficient, carrier density, and electrical resistivity, instead of the exponential dependence that characterizes a gapped semiconducting state.\cite{vaipolin92} Fig.~\ref{fig4} displays the electrical resistivity plotted as a function of temperature on a $\log$-$\log$ scale in order to assess the possibility of power law behavior. At ambient pressure (\textit{black squares}), a linear region extends from room temperature down to 30 K and corresponds to a power law exponent $n \sim -1$. As pressure is increased, the linear regions in the $\log$-$\log$ plots extend over a smaller temperature range, preventing an accurate power-law fit at higher pressures. The power-law behavior observed at ambient pressure is consistent with the zero-gap semiconducting behavior previously suggested in Refs.~\onlinecite{vaipolin92,popov11}, although in those works, a somewhat different power of $n \sim -1.9$ was found. One factor that may complicate the above analysis of the temperature dependence of the resistivity is that the pressure in the hyrdrostatic cell is not perfectly constant between low temperature (where the pressure is measured) and room temperature. This is largely due to the freezing, and associated decrease in volume, of the n-pentane-isoamyl alcohol pressure medium. Previous experience with similar pressure cells indicates that the increase in pressure from low temperature to room temperature could be as large as 0.25 GPa.\cite{wohlleben_1971_1}

The results of the diamond-anvil cell experiments at temperatures from room temperature to 1.2 K are presented in Fig.~\ref{fig5}. Red, green and blue data were collected during the cooling cycles for pressures of 4, 14 and 34 GPa, respectively, while the black curves correspond to the measurements taken during warm-up of the probe. The upper panel of Fig.~\ref{fig5} displays the electrical resistance curves, showing that at 34 GPa the resistance begins to decrease with decreasing temperature below 150 K. This effect can be observed more clearly in the lower panel of Fig.~\ref{fig5}, which shows plots of resistance, normalized to the value at 270 K. A small but sharp drop of the resistance is evident at $\sim$ 8 K and 34 GPa.  The $inset$ of the lower panel of Fig.~\ref{fig5} displays the low temperature region of the 34 GPa curve, measured during warm-up. This sharp drop is most probably due to superconductivity arising from inclusions of elemental tellurium, which becomes metallic near 5 GPa and has a high-pressure superconducting phase above 30 GPa.\cite{akahama92} At 4 GPa (upper panel $inset$), a band gap of approximately 3 meV can be obtained by fitting the high temperature resistance. This energy gap value is consistent with the pressure dependence of the energy gap on pressure presented in Figure~\ref{fig3} for the hydrostatic cell experiment, which suggests that the energy gap is almost closed at 4 GPa.  As for the hydrostatic cell, one complicating factor in this analysis of the temperature dependence of the resistivity is that the pressure may change between low temperature and room temperature. In contrast to the hydrostatic cell, where pressure drops somewhat with cooling, our experience with this type of diamond anvil cell suggests that the pressure may increase by 10-20\% upon cooling.

\begin{figure}[t]
	\begin{center}
	{\includegraphics[width=3in]{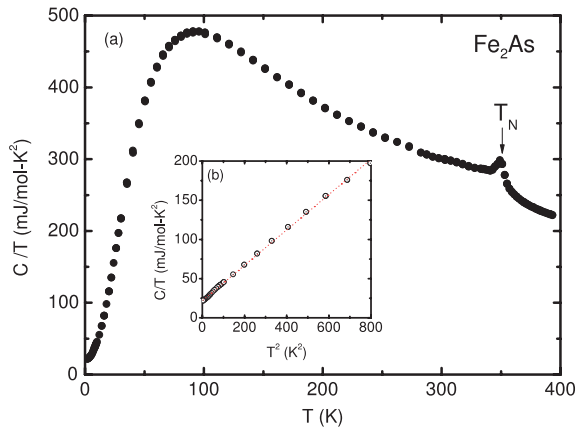}}
	\caption{(a) Specific heat $C$, divided by temperature $T$, versus $T$ for Fe$_2$As. (b) $C/T$ versus $T^2$ and the linear fit at low temperatures (see text).}
	\label{fig6}
	\end{center}
\end{figure}

Given that the energy gap appears to close at a pressure near 4 GPa, it is interesting that the resistance continues to increase with decreasing temperature to significantly higher pressures of at least 14 GPa.  Such behavior might be explained by localization due to the substantial amount of disorder associated with the mixed occupation and vacancies in the Cu/Fe sites.  This possibility is supported by the relatively high electronic density (10$^{22}$ cm$^{-3}$) and low mobility (10$^{-1}$ cm$^{2}$V$^{-1}$s$^{-1}$) at room temperature and ambient pressure previously reported for CuFeTe$_2$.\cite{vaipolin92} However, the resistance is not described, at any pressure, by three dimensional variable range hopping behavior $R(T) = R_0\exp[(T_0/T)^{1/4}]$.  A detailed study of the pressure dependence of the Hall coefficient and the carrier concentration could be useful in exploring these possibilities. It remains unclear whether CuFeTe$_2$ at ambient pressure belongs to the family of zero-gap semiconducting materials, such as graphene or tellurium-based HgTe and CdTe, since these materials generally display power-law temperature dependences of their transport properties, similar to CuFeTe$_2$ (at ambient pressure), but they are also characterized by low carrier concentrations and extremely high mobilities.\cite{wang2010}

The main panel of Fig.~\ref{fig6} presents specific heat $C(T)$ data for Fe$_2$As in the range 2-400 K.  A sharp feature at 350 K appears at the N\'{e}el temperature.  Shown in the inset is a plot of $C/T$ versus $T^2$, where the dashed red line represents a linear fit to the data with a $T=0$ K intercept $\gamma$ and slope $\beta$.  Here, $\gamma$ and $\beta$ are, respectively, the electronic and phonon contributions to the low temperature specific heat, which can be described by the expression $C(T) = \gamma T + \beta T^3$.  The fit yields the values $\gamma = 21.8$ mJ/mol-K$^2$ and $\beta = 0.23$ mJ/mol-K$^4$, where the value of $\beta$ corresponds to a Debye temperature $\Theta_D = 296$ K.  The moderately enhanced value of $\gamma$ is likely due to a peak in the density of states at the Fermi level associated with the Fe $3d$-electrons.\cite{chonan_1991_1}

\begin{figure}
	\begin{center}
	{\includegraphics[width=3.3in]{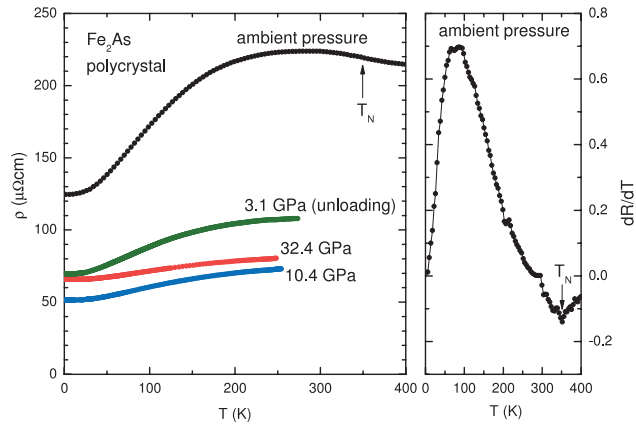}}
	\caption{Electrical resistivity of Fe$_2$As at ambient and high pressure.}
	\label{fig7}
	\end{center}
\end{figure}

Figure~\ref{fig7} shows electrical resistivity versus temperature for Fe$_2$As at both ambient and high pressures. In the ambient pressure data, a small feature is visible in the electrical resistivity near $T_N$. The change in sign of the resistivity slope near room temperature may also be related to the antiferromagnetic ordering. Upon increasing pressure, the overall shape of the resistivity remains similar to the ambient pressure curve. The resistance does appear to drop with pressure, although uncertainties in the geometric factor used to convert resistance to resistivity could produce an error of up to 50\% in the absolute resistivity values for the high pressure data. From the present data, it is not possible to determine the evolution of the N\'{e}el temperature with pressure.  Further measurements, above room temperature, would be required to map the initial pressure dependence of $T_N$.

\section{Summary}
In summary, we have synthesized samples of Fe$_2$As and CuFeTe$_2$. Both materials remain non-superconducting above 1.1 K to pressures above 30 GPa. At ambient pressure, the CuFeTe$_2$ samples display non-metallic behavior. The high-pressure electrical resistivity measurements performed on single crystals of CuFeTe$_2$ indicate that a modest applied pressure of 2.4 GPa is enough to decrease the values of electrical resistivity at 4 K by two orders of magnitude, and an energy band gap seems to be suppressed near 4 GPa. At 34 GPa, the resistivity decreases upon cooling below 150 K, suggesting the possibility that CuFeTe$_2$ has been driven metallic. An important unanswered question, which we plan to address in the future, is whether these materials remain in the Cu$_2$Sb structure to high pressures or undergo transformations to other structure types.

\section{Acknowledgements}
High-pressure research at University of California, San Diego, was supported by the National Nuclear Security Administration under the Stewardship Science Academic Alliance program through the U.S. Department of Energy grant number DE-52-09NA29459.  Sample synthesis was supported by AFOSR-MURI, Grant FA9550-09-1-0603, while physical properties characterization at ambient pressure was supported by DOE Grant DE-FG02-04-ER46105.  Lawrence Livermore National Laboratory is operated by Lawrence Livermore National Security, LLC, for the US Department of Energy (DOE), National Nuclear Security Administration (NNSA), under Contract No. DE-AC52-07NA27344. Y.K.V.\ acknowledges support from DOE-NNSA Grant No. DE-FG52-10NA29660. D.Y.T.\ thanks The Scientific and Technological Research Council
of Turkey (T\"{U}B\.{I}TAK) for partial support.

\bibliography{hamlin,zocco}

\end{document}